\documentclass[aps,pra,twocolumn,groupedaddress]{revtex4-1}
\usepackage[latin1]{inputenc}
\usepackage[T1]{fontenc}
\usepackage{bm}
\usepackage{graphicx}
\usepackage{amsmath}
\usepackage{amsfonts}
\usepackage{amssymb}
\usepackage{color}
\usepackage{epstopdf}
\DeclareGraphicsExtensions{.pdf,.eps,.png,.jpg,.mps} 
\begin{document}

\title 
{Electrical control of spin relaxation time in complex  quantum nanostructures.}

\author{M. Kurpas}
\author{B. K\k{e}dzierska}
\author{I. Janus--Zygmunt}
\author{M. M. Ma\'{s}ka}
\author{E. Zipper} 
\affiliation{Institute of Physics, University of Silesia ul. Uniwersytecka 4, 40-007 Katowice, Poland }

\begin{abstract}
Spin related phenomena  in quantum nanostructures  have attracted  recently much interest due to fast growing field of spintronics. In particular complex nanostructures are important as they provide a versatile system to manipulate spin and the electronic states. Such systems can be used as spin memory devices or scalable quantum bits. We investigate the spin relaxation for an electron in a complex structure composed of a quantum dot surrounded by a quantum ring. We shown that modifications of the confinement potential result in the substantial increase of the spin relaxation time. 
 
\end{abstract}

\maketitle
\section{Introduction}\label{sec1}

Quantum nanostructures (QNs) exhibit new physics which have no analogue in real atoms. One of the reasons is the nature of the confining potential which is different than the $1/r$ potential of atoms and which can be tuned to desired shapes and heights.
Besides, by changing the gate voltage of the  device one can change the number of electrons down to zero one by one \cite{hans}.
Due to large dimensions (comparing to atoms) of quantum dots (QDs) or quantum rings (QRs) and because they can be integrated in electrical circuits, they are suitable for experiments that cannot be carried out with normal atoms. In particular QNs can be used as spin memory devices or quantum bits in quantum computing \cite{vander}.\\
Modern nanotechnology allows fabrication, depending on the destination, various complex nanostructures in which some characteristics can be optimized. The  potentials confining electrons in nanostructures depend on the method of fabrication and the external voltages applied to the leads  e.g. etched nanostructures have deeper potential comparing to electrostatically defined ones.  
In the theoretical analysis one assumes certain model potential profiles e.g. the harmonic one \cite{Khaetskii}. However experimental results suggest that the real confinement potential is non-parabolic and usually posses well-like structure \cite{bayer}.

The answer to the question how important the confinement potential is for QDs, was roughly given by comparing energy spectra of 2D quantum harmonic oscillator and infinite square well. The energy levels of the former model are equidistant while for the latter one they scale as $n^2$ where $n$ numbers the energy states.

In this paper we discuss a two-dimensional complex structure composed of a quantum ring with the quantum dot inside (dot--ring nanostructure, DRN).
We want to check how the confinement potential details influence sample characteristics of the DRN, in particular we will focus on spin relaxation.
This problem has been already addressed in \cite{wfe} for the confinement potential of the Gaussian type. We studied there the high controllability of the DRN by demonstrating  how its coherent and optical properties can be changed by modifications of the Gaussian potential. 
In this paper we want to examine the influence of a broad class of realistic potentials from almost rectangular to 
the potential with smooth edges on spin relaxation.
 
 
To make the analysis transparent but still to take advantage of the complexity of the system  we consider the DRN to work in the Coulomb blockade regime with a {\it single} electron. 
The energy cost for adding an extra electron to the DRN is large enough that one can adjust the parameters of the device allowing at most one electron to occupy the DRN \cite{hans}.


The paper is organized as follows: In Sec. \ref{sec2} we present a general theoretical background 
for studying quantum coherent properties of DRNs used as spin memory devices or quantum bits. In Sec. \ref{sec3} we 
 demonstrate how we can control spin relaxation times by changing the shape of the confinement potential. 
The results are summarized in Sec. \ref{sec4}.

\section{Quantum confinement of dot-ring nanostructure}\label{sec2}

\begin{figure}[h]
\begin{center}
\includegraphics[width=0.9\linewidth]{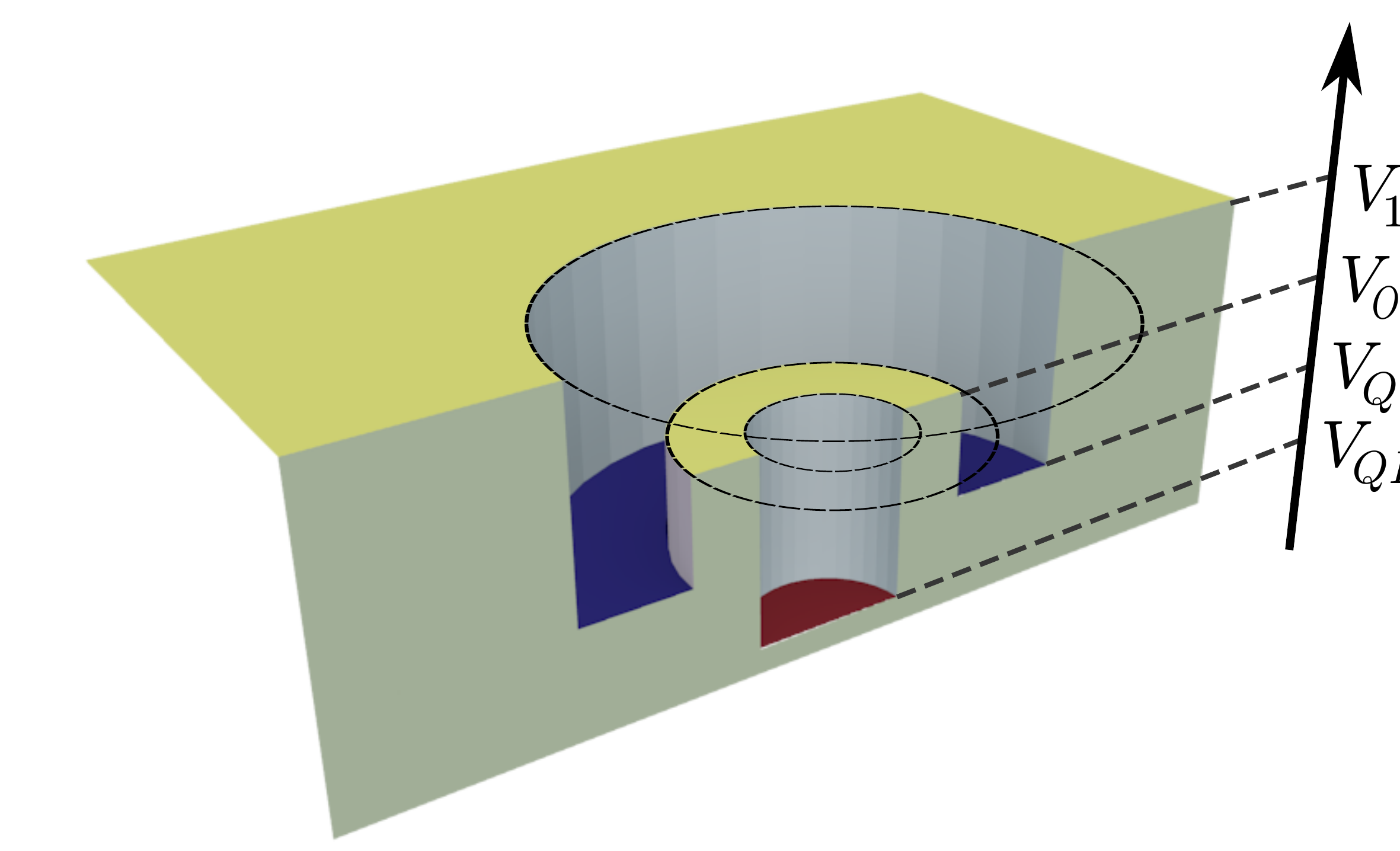}\vspace*{2mm}
\caption{Cross section of the rectangular confinement potential forming the DRN. The bottom of the ring's and dot's part of the nanostructure are marked with blue and red color respectively. $V_1$ is the height of the potential at the outer edge of the DRN, $V_{QD}$ and $V_{QR}$ are bottoms of the potentials of the dot and ring respectively and $V_0$ is the split--barrier height.}
\label{fig1}
\end{center}
\end{figure}

We consider a two dimensional circularly symmetric QN defined by a confining potential $V(r)$ obtained by introducing into a circular quantum dot (QD) a circular split--barrier $V_0(r)$ \cite{peeters} that divides the initial quantum dot into a quantum ring with the quantum dot inside (DRN) (Fig. \ref{fig1}). We place the nanostrucutre in a static magnetic field $B$ parallel to its plane and assume the nanostructure, in particular the barrier $V_0$, to be controllable e.g. by electrical gating. 
For illustration a cross section of a rectangular potential with explanations of symbols used throughout the text is presented in Fig. \ref{fig1}. Such a structure with the confinement potential which conserves the circular symmetry (and therefore the orbital degree of freedom is a good quantum number) can be fabricated e.g. by the split gates technique \cite{zhitenev} or by pulsed droplet epitaxy \cite{somaschini}.

The single--electron Hamiltonian is written as

\begin{equation}
H = \frac{{\bf p}^2}{2m^{*}}  + \frac{e\hbar}{2m^*} \mathbf{\hat{\sigma}} \cdot \mathbf{B}+V(r), 
\label{Hamiltonian_pr}
\end{equation} 
where $ m^{*}$ is the effective electron mass, $V(r)$ is the potential defining the structure.\\
For concreteness, our model calculations are performed for InGaAs systems (with $m*= 0.067m_e$, $g_s=|0.8|$) for which many of the 
theoretical and experimental investigations have been performed.\\
The single electron energy spectrum of $ H$ consists of a set of discrete states $E_{nl}$ due to radial motion with radial quantum numbers $n=0,1,2,\ldots$, and rotational motion with angular momentum quantum numbers $l=0,\pm 1,\pm 2\ldots$.
The single particle wave functions are of the form 
\begin{equation}
\Psi_{nl\sigma} = R_{nl}\left(r\right)exp\left(i l \phi \right)\chi_{\sigma},
\label{eq_psi_nl}
\end{equation}
with the radial part $R_{nl}(r)$ and the spin part $\chi_{\sigma}$.

We can then  calculate the so called 'overlap factor' (OF) which depends on the distribution of 
the wave functions in the DRN.
It is given by:
\begin{equation}
\Xi_{n'l',nl} = \int_0^{\infty} R_{n'l'}^{*}R_{nl}r^2 dr,
\label{xi}
\end{equation}
where $(n'l'), (nl)$ are the quantum numbers of the two energy states involved in the process under investigation.\\
In the following we consider a DRN occupied by a {\it single} electron coupled to phononic degrees of freedom so the relevant OF is
\begin{equation}
\Xi_{00,nl}\equiv \Xi_{nl} = \int_0^{\infty} R_{00}^{*}R_{nl}r^2 dr,
\label{xi0}
\end{equation}
where $ R_{00}$ is the radial part of the ground state wave function (\ref{eq_psi_nl}) and $l=\pm 1$, due to the selection rule for phonon transitions.

The second important quantity is the energy gap between the orbital excited and the ground state 
 $\Delta_{nl}=E_{nl}-E_{00}$. 
The numerically calculated  energy spectra, modified by electrical gating, allow us to estimate relaxation time for a set of DRNs.

\begin{figure}[h]
\begin{center}
\hspace*{-5mm}\includegraphics[width=\linewidth]{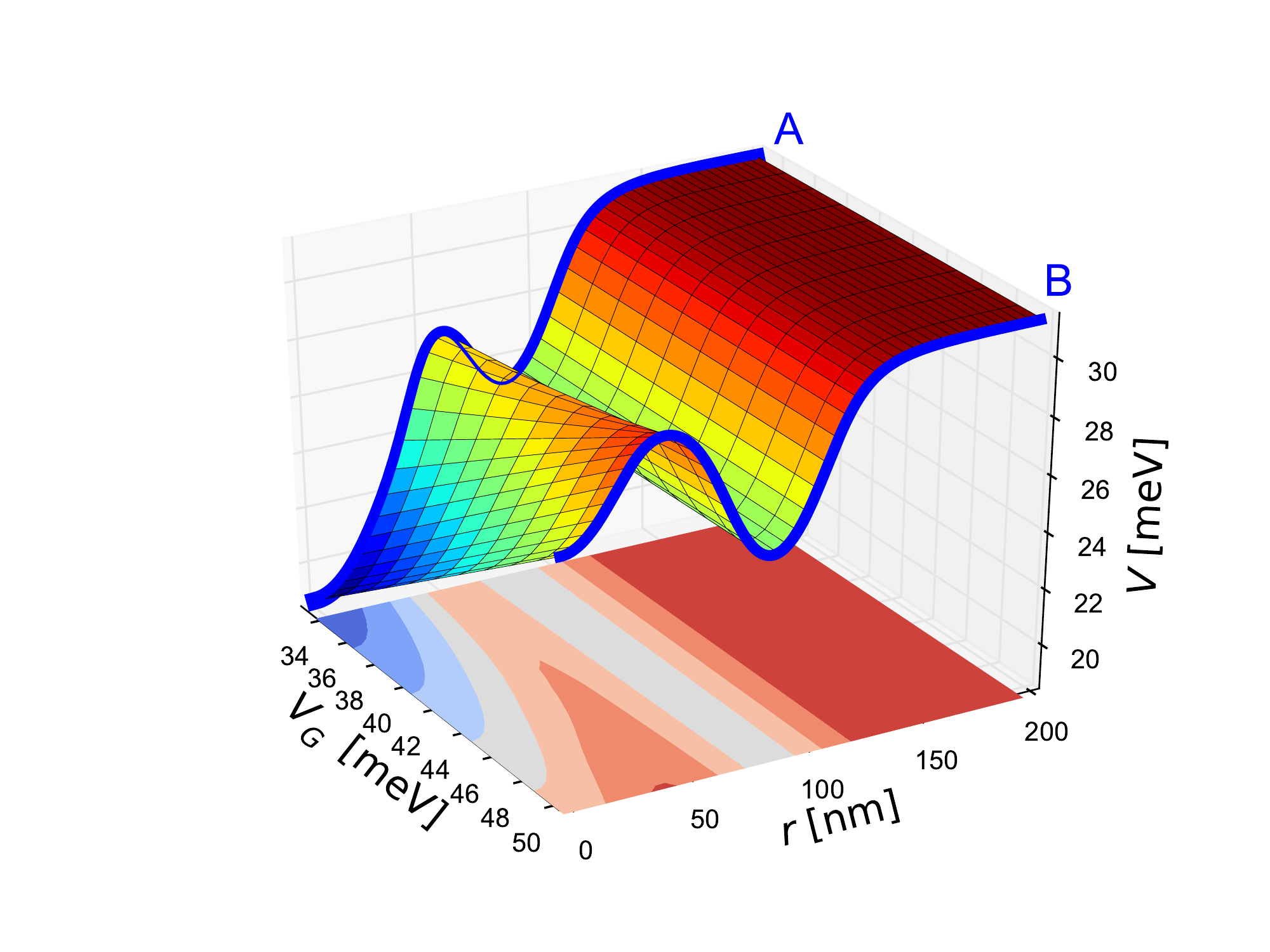}\vspace*{2mm}
\caption{ Cross section of the confining potential obtained from solving the Laplace equation. $V_G$ is the voltage of the top-gate forming the dot part of the structure (the ring's top-gate voltage is kept constant). Lines A and B correspond to solutions with the ground state is in the QD and QR, respectively.}
\label{fig_poisson}
\end{center}
\end{figure}

At first we consider an elctrostatically defined DRN \cite{peeters} in which the potential confining electrons is generated by two planar, concentric electrodes -- a circular one in the center surrounded by a ring shaped electrode. 
For such a system one can solve the Laplace equation to get the profile of the potential felt by the electrons in 2DEG. Changing the voltage applied to the electrodes one can obtain a diverse class of confinement potentials. Fig. \ref{fig_poisson} demonstrates the evolution of the shape of the potential as the voltage applied to the dot's part electrode is varied while the voltage of the ring's electrode is kept constant. 

The spin of a {\it single} electron in circularly symmetric QN placed in a static magnetic field $B$ with energy levels split by the Zeeman energy
 $\Delta _Z=g_s \mu_B B$ provides a natural system suitable as memory device in spintronics and as a qubit in a quantum computer \cite{vander}. If $k_BT\ll\Delta _Z\ll\Delta_{01}$  then, the QN can be well approximated as a two level system. In our model calculations, we also assume the in-plane magnetic field $B=1.5T$ (the influence of magnetic field on the orbital states is then very small)  thus, the Zeeman splitting is equal to $\Delta_Z=0.069$\:meV.

The DRN qubit can be initialized, e.g., by thermal equilibration or by optical pumping, coherently manipulated 
(through magnetic resonance technique or by faster electrical and optical gates) and read out using both electrical
and optical techniques \cite{elz,hans,mano}. Coherent coupling of electrostatically defined DRNs, leading to the formation of, e.g., the CNOT gate can be obtained in an analogous way as for QDs \cite{vander}: one can assemble a system of two coplanar DRNs with the
 possibility of tuning their exchange coupling $J$ by gating the barrier between them. 
 Quantum gates for self-assembled DRNs can be implemented by electronic or photonic connections \cite{mano,ladd,Abba}.
 Single qubit rotations together with the CNOT gate form a universal set of quantum logic gates. Remarkably, these operations are very fast, of the order of pico to nanoseconds \cite{przegladLossa}. We discuss below how to optimize material properties and nanostructure design to achieve long relaxation times for spin memory device and both relaxation and decoherence times for spin qubits, so that a sufficient room is left for implementing protocols for spin manipulations and read out. 
Then many coherent operations can be performed during relaxation and decoherence.
\section{Spin relaxation in complex nanostructures}
\label{sec3}

For magnetic field $0.1 < B< 10$T spin relaxation rates are given by single-phonon emission processes accompanied by spin flips arising from spin-orbit coupling \cite{Khaetskii,golo}. 
The rates for piezoelectric phonon coupling in the discussed materials are greater than those from deformation potential. 
It was shown theoretically \cite{Khaetskii,stano,golo} and confirmed experimentally \cite{kroutvar, amasha, heiss}.

The formula for the relaxation time $T_1$ governed by the Dresselhaus spin--orbit (SO) interaction is given by (for detailed derivation see \cite{Khaetskii}):
 
\begin{equation}
\frac{1}{T_1}=\frac{\Delta_Z^5}{\eta}\left[ \sum_{n,l} \frac{\Xi_{nl}^2}{\Delta_{nl}}\right]^2
\label{t1_r}
\end{equation}

\begin{equation}
\eta=\frac{\hbar^5 }{\Lambda_p (2\pi)^4 (m^{*})^2 },
\end{equation}

$\Lambda_p$ is the dimensionless constant depending on the strength of the effective spin-piezoelectric phonon coupling and the magnitude of SO, $\Lambda_p=0.007$ for GaAs type systems \cite{kroutvar,Khaetskii}.

The extensive discussion of relaxation for quantum rings has been  given in \cite{zkm}. It was shown there that $T_1$ increases significantly with the decrease of the radius $R$ of the structure. We also found this effect to occur for the DRN. For concreteness we present the results for $R=90$ nm.
 
It follows from our calculations  that the relaxation time for DRN is determined by the SO coupling to the two lowest excited orbital levels, thus 
 \begin{equation}
\frac{1}{T_1}=\frac{\Delta_z^5}{\eta}\left[ \Gamma^{01} + \Gamma^{11}\right]^2,
\label{t1_dr}
\end{equation}
where $\Gamma^{01}=\Xi^2_{01}/\Delta_{01}$, $\Gamma^{11}=\Xi^2_{11}/\Delta_{11}$.

The quantities entering $T_1$ depend on  the orbital energy spectra and the overlap factor $\Xi$, which are strongly related to the shape and the distribution of the orbital wave functions. These two parameters are determined by the confinement potential of the structure. 

\begin{figure}
\includegraphics[width=\linewidth]{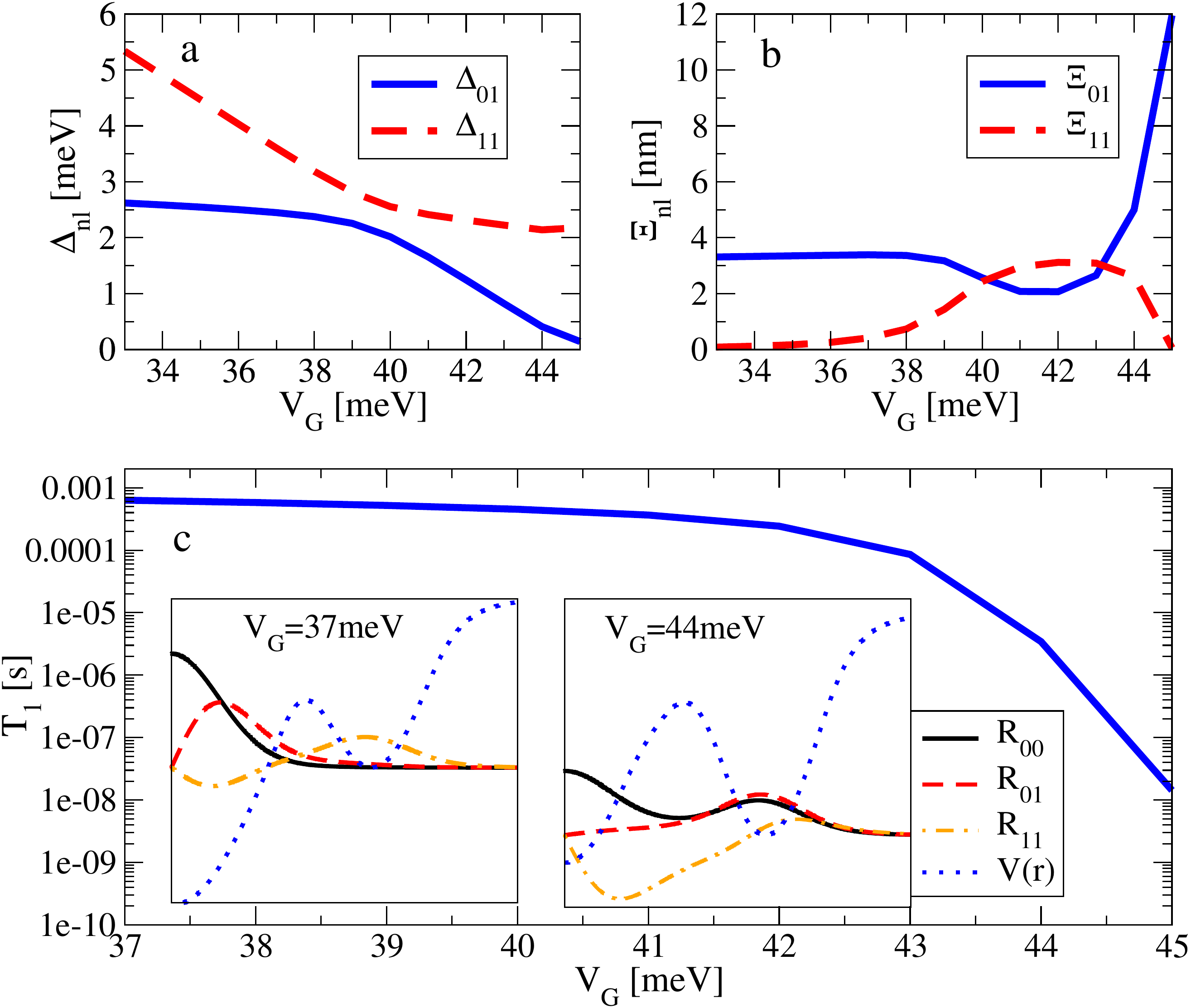}
\caption{(color online) Dependence of the orbital energy gaps (a), overlap factor (b) and relaxation time (c) on the dot's part top-gate voltage $V_G$. The potential $V(r)$ entering Eq. (\ref{Hamiltonian_pr}) has been taken from  solving the Laplace equation for each value of $V_G$. In the insets the distribution of the wave functions for two cases is shown.}
\label{3plot_VG}
\end{figure}

At first let us focus on the potential obtained from the Laplace equation (shown in Fig. \ref{fig_poisson}). By changing $V_G$ one can change the level of the QD's potential minimum keeping QR's potential unchanged (to some extend).
This affects the distribution of the wave functions in the structure and hence the relaxation (Fig. \ref{3plot_VG}). We see that increasing $V_G$ (ground and excited state wave function move to QR) the relaxation time decreases. This is caused by stronger (comparing to QD) overlap of the wave functions in QR and smaller energy level quantization for the lowest states (Fig. \ref{3plot_VG}a,b).

\begin{figure}
\includegraphics[width=\linewidth]{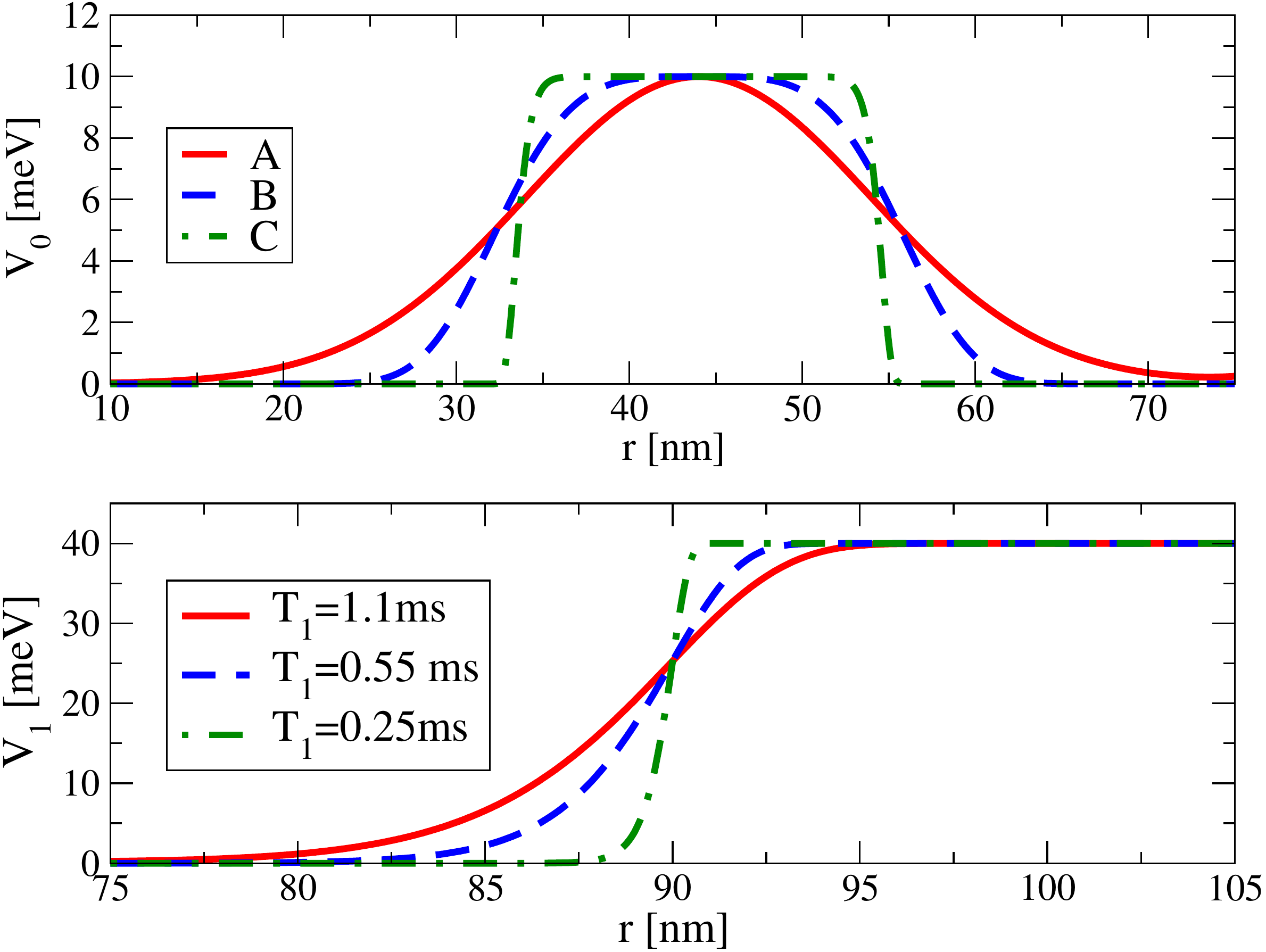}
\caption{(color online) Differences in confinement shape for three potentials A,B and C with corresponding relaxation times $T_1$. Top pane: the shape the potential barrier $V_0$, bottom pane: the slope of the outer edge ($V_1$) of the DRN. Both parts of the potential were changed simultaneously but are plotted separately for clarity; $V_0=10$meV, $V_1=40$meV, $V_{QD}=V_{QR}=0$.}
\label{f_bump_shape}
\end{figure}

Now we will consider other potential shapes. QNs fabricated within modern nanotechnology are produced by different methods (pulsed droplet epitaxy\cite{somaschini}, electrostatically defined QNs \cite{peeters,kroutvar, amasha, elz}, deep-etched structures, modulated barrier structures \cite{bayer}) equivalent to varying lateral confinement potential \cite{bayer}. Besides, plunger gates can continuously deform the confinement.
The confinement was found to have significant influence on various QD characteristics such as conductivity, weak localization corrections \cite{ouch}, exciton binding energy \cite{bayer}.
Also, in the case of the DRN one should expect a similar effect.  In Fig. \ref{f_bump_shape} we present three different shapes of the confinement for the same values of $V_{QD},V_{QR},V_0$ and $V_1$. The calculated relaxation times are given in the inset. 
It turns out that the more rectangular the potential is the faster is the relaxation. This result can be understood in terms of size quantization: smooth potential (solid red line) spreads over almost whole nanostructure squeezing the wave function much more than a step--like one (dash-dot green line) that is more narrow and the effective size of the structure is bigger.

\begin{figure}
\includegraphics[width=\linewidth]{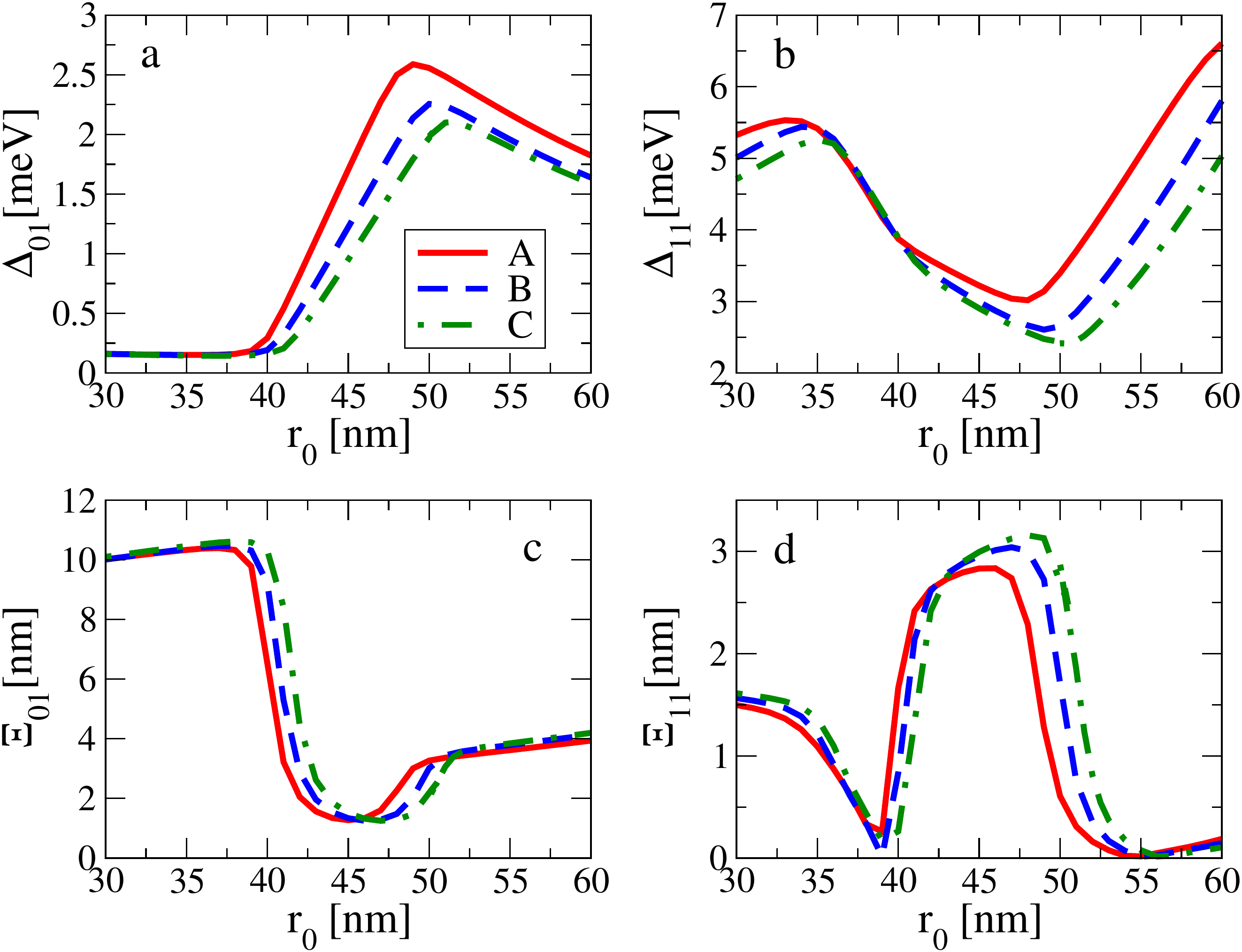}
\caption{(color online) Dependence of the orbital energy gaps (panes a, b) and overlap factors (panes c,d) on the position of the split--barrier $V_0$ plotted for three different potential shapes.}
\label{f_delta_ksi_br0}
\end{figure}

\begin{figure}
\includegraphics[width=\linewidth]{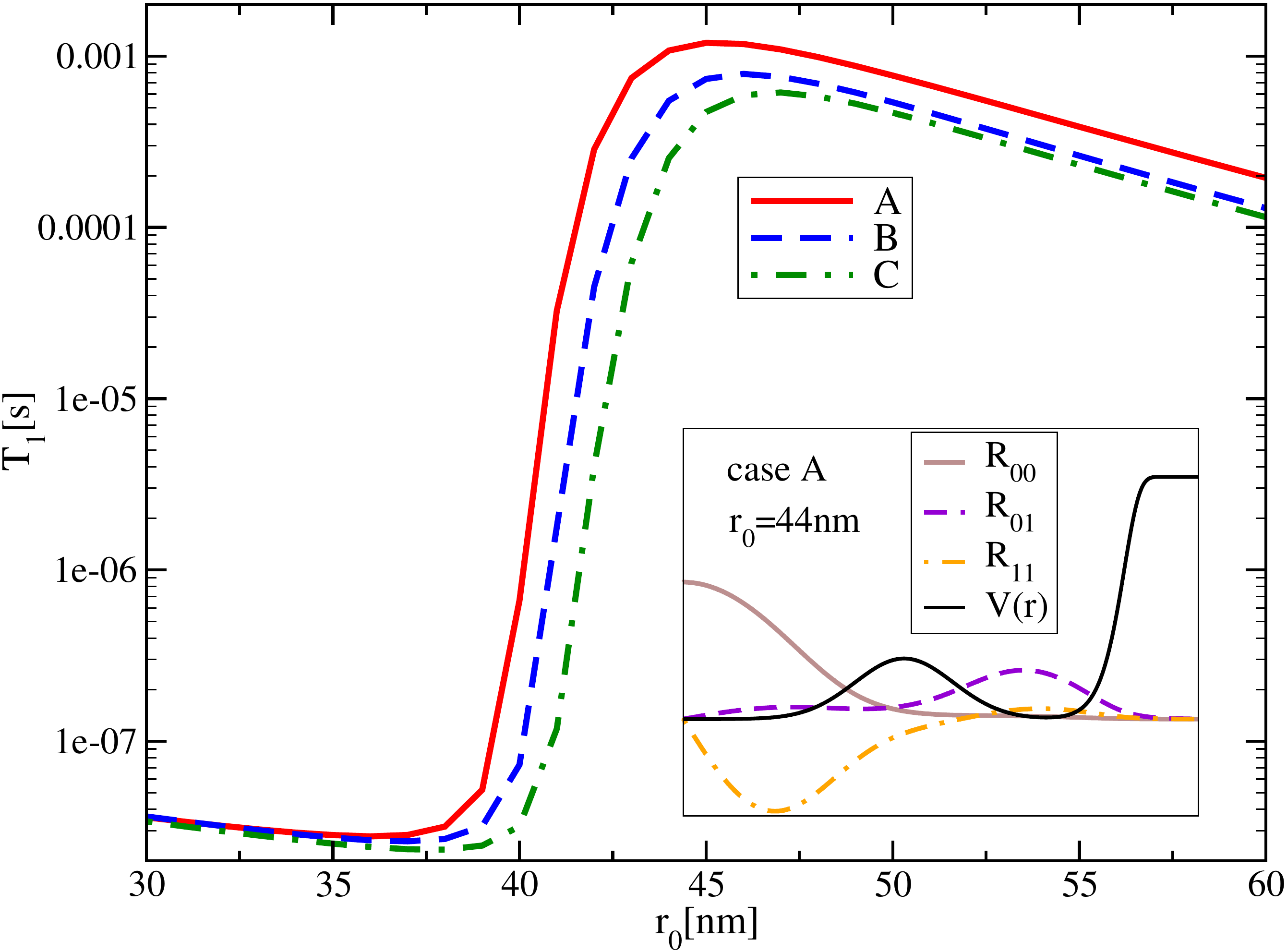}
\caption{(color online) Dependence of the relaxation time $T_1$ on the position of the split--barrier $V_0$ plotted for three different potentials A,B,C. In the inset the distribution of the radial parts of the wave functions for $r_0=44$nm for the potential A is shown.}
\label{f_T1_br0}
\end{figure}

It is also not straightforward to guess what is the optimal relation between the size of the dot and the ring composing the DRN that maximizes $T_1$. To answer this question we studied the case where the position $r_0$ of the barrier $V_0$ was varied and the remaining parameters were kept constant. As $T_1$ is expressed via $\Delta_{nl}$ and $\Xi_{nl}$ at first we investigate how they depend on $r_0$. The results shown in Fig. \ref{f_delta_ksi_br0}a--d allows us to expect strong dependence of $T_1$ on $r_0$.
If the QD is small (Fig. \ref{f_T1_br0}, $r_0<40$\:nm) it is then energetically favorable for the system to move the wave functions to the ring which is thick. 
Moving the barrier towards the outer edge of the DRN [the width of QD (QR) increases (decreases)] relaxation time increases and has a maximum when the barrier is approximately in the middle of the DRN (see inset in Fig. \ref{f_T1_br0}). Further increase of $r_0$ does not affect $T_1$ very much. 

Thus by manipulating the orbital states of QN using gate voltages, one can control the amount of spin--orbit--induced mixing of different excited states and vary $T_1$ by over an order of magnitude.\\%
One should stress that the important feature of such studies is not the value of $T_1$ itself but the possibility to control it by external conditions which can be steered by electric fields. 

These long relaxation times have been obtained taking into account only SO mediated interaction with piezoelectric phonons which is the dominant mechanism of spin relaxation for nanostructres made of III-V semiconductors and of relatively small size \cite{Khaetskii}.
Considering also other mechanisms of relaxation, (e.g., due to fluctuations of the electric and magnetic field, deformational phonons, multiphonon processes, and circuit noise) in the above model calculations, can further limit the relaxation time. However, it was shown that in the considered  parameter regime they are substantially smaller \cite{Khaetskii,golo}.

The spin decoherence time $T_{2}$ for nanosystems made out of III-V semiconductors is limited by hyperfine interaction (HFI) as the SO mechanism does not lead to pure dephasing \cite{golo}. 
Recent experiments have shown that  $T_{2} \sim 10 - 100\:\mu s$ \cite{yao,bluhm} for the considered magnetic field. Besides, several possibilities have been proposed to decrease the randomness in the nuclear-spin system:
 polarization of nuclei \cite{xu} and putting the nuclear spins in a particular quantum state \cite{Giedke} seem to be very promising. 

Because of the destructive role  of nuclear spins on the coherence of electron spins one can use different material. If DRNs were made not of III-V semiconductors (with non-zero
nuclear spin) but of the group IV isotopes with zero nuclear spins, the coherence times should be longer because of the absence or very small hyperfine interaction.
 As a result one could then get $T_{2} = 2T_{1}$, which is a relatively long time.

Because of the significant role of silicon in modern electronics the estimations for silicon--based nanostructures \cite{stano1} are important. 
It is known that the magnitude of the spin--orbit coupling in silicon is ten times smaller than in GaAs. Besides, for silicon the deformation phonon potential is the most important due to  the absence of piezoelectric phonons. All this  results in the relaxation times hundred times longer. 
At the same time these systems should have long decoherence times $T_{2} = 2T_{1}$ due to the absence of nuclear spins. 
Thus silicon, the important semiconducting material for charge based electronics also seems to be an important choice for spin based electronics  and for quantum computing \cite{aws}. 
Summarizing, the decoherence time of electron spins in material with few or no nuclear spins 
 are expected to be longer than for the group III-V semiconductors. 
However, in all considered materials the decoherence times are much longer than the initialization, qubit operations and measurement times allowing for quantum error correction scheme to be efficiently used \cite{przegladLossa}.

\section{Summary}\label{sec4}
Quantum information processing and spintronics have been  the major driving forces towards the full control of single--spin systems.
Deep understanding of underlying physics allows to propose quantum systems in which some properties can be optimized. In particular we have looked for the QNs with long spin relaxation time. 
We performed systematic studies of the influence of the shape of the confinement potential on relaxation. We investigated different confining potentials which correspond to different methods of fabrication. 
It follows from the presented considerations that relaxation depends crucially on the arrangement of the electron wave functions.
Both the energy spectrum and overlap factors are sensitive functions of the confining potential and various structures with desired properties can be engineered.
 
In particular, we have shown how the evolution of the confinement potential shape from the nearly rectangular to the potentials with smooth boundaries influences the spin relaxation.
When we move over from the rectangular potential to the potential with smooth edges the effective radius of DRN decreases which results in an increase of $T_1$. This is consistent with the finding that $T_1$ is longer for smaller radius of DRN.
We have also found that relaxation depends significantly on the position of the split--gate and that the optimal  position can be found to maximize relaxation time.
  
Our studies confirm that combined quantum structures are highly relevant to new technologies in which the control and manipulations of electron spin and wave functions play an important role. Thus quantum carrier confinement in circular nanostructures can be the basis of many applications in spintronics and quantum computing devices.
Our results should serve as a hint for experimentalists in order to fabricate QNs with, depending on destination, the best properties. 

\acknowledgments
One of the authors (M.M.M.) acknowledges support from the Foundation for Polish Science under the TEAM program for the years 2011--14. The authors thank M. Nowak and B. Szarfan for providing the code that was used in the calculations of the Laplace equation.

\end{document}